\documentclass[letterpaper,12pt]{article}

\pdfoutput=1
\usepackage{graphicx}
\usepackage{amsmath}
\usepackage{amsfonts}
\usepackage{hyperref}
\usepackage{bm}

\makeatletter
\renewcommand\section{\@startsection {section}{1}{\z@}%
{-3.5ex \@plus -1ex \@minus -0.2ex}%
{2.3ex \@plus 0.2ex}%
{\normalfont\normalsize\bfseries}}

\renewcommand\subsection{\@startsection{subsection}{2}{\z@}%
{-3.25ex \@plus -1ex \@minus -0.2ex}%
{1.5ex \@plus 0.2ex}%
{\normalfont\normalsize\bfseries}}

\def\@seccntformat#1{\csname the#1\endcsname.\quad}
\makeatother

\begin{document}

\setlength{\baselineskip}{3.75ex}

\noindent
\textbf{\LARGE The fiducial-Bayes fusion}

\vspace{2ex}
\noindent
\textbf{\Large A general theory of statistical inference}

\vspace{7ex}
\noindent
\textbf{Russell J. Bowater}\\
\emph{Independent researcher, Doblado 110, Col.\ Centro, City of Oaxaca, C.P.\ 68000, Mexico.\\
Email address: as given on arXiv.org. Twitter profile:
\href{https://twitter.com/naked_statist}{@naked\_statist}\\ Personal website:
\href{https://sites.google.com/site/bowaterfospage}{sites.google.com/site/bowaterfospage}}

\vspace{5.5ex}
\noindent
\textbf{\small Abstract:}
{\small An overview is presented of a general theory of statistical inference that is referred to
as the fiducial-Bayes fusion. This theory combines organic fiducial inference and Bayesian
inference. The aim is that the reader is given a clear summary of the conceptual framework of the
fiducial-Bayes fusion as well as pointers to further reading about its more technical aspects.
Particular attention is paid to the issue of how much importance should be attached to the role of
Bayesian inference within this framework.
The appendix contains a substantive example of the application of the theory of the fiducial-Bayes
fusion, which supplements various other examples of the application of this theory that are
referenced in the paper.}

\vspace{2.5ex}

\vspace{4ex}
\section{Introduction}

The aim of this paper is to present a guide to a general theory of statistical inference that will
be called the fiducial-Bayes fusion, which combines organic fiducial inference as developed in
Bowater~(2018, 2019, 2021) and Bayesian inference.

We will be concerned with the broad issue of making inferences about the parameters
$\theta = \{\theta_i: i=1,2,\ldots,k\}$ of a sampling distribution on the basis of an observed data
set $x$ randomly drawn from this distribution when it is known that the distribution concerned
belongs to a given parametric family of distributions.
This issue is important both in its own right, and because tackling this issue can be viewed as a
natural stepping stone to resolving the problem of what to do when it is inadequate to assume that
the true sampling distribution belongs to a given family of distributions.
To clarify, the form of the sampling distribution may be dependent on additional observable
variables, e.g.\ what would be termed independent variables in a regression analysis, and therefore
here we are using the term `sampling distribution' to encompass what is often referred to as a
`statistical model'.

\vspace{3ex}
\section{Statistics by analogy}

Making analogies is a fundamental part of performing statistical inference. Indeed, arguably the
most natural way of making sense of the concept of probability is through the use of analogy. Some
of these analogies may be good, some adequate and some poor. In many situations, there is a variety
of competing analogies that we can make and choosing to use one of these analogies rather than
another can radically affect what method of inference will be implemented, and as a consequence,
can have a substantial impact on what conclusions will be drawn on the basis of the observed data.
It should be obvious without the need for further elaboration that we should always want to make
the best analogies as possible. The mathematical tradition in the field of statistics has
overlooked the importance of the use of analogy in making inferences, which has led to much
inconsistency and many flaws in statistical theory.

\vspace{3ex}
\section{Probabilistic inference}

Under the restriction that the sampling distribution of the data $x$ belongs to a given parametric
family of distributions, we will assume that the ultimate goal of statistical inference is as
follows:
\par
\begin{quote}
To place a probability distribution over the parameters $\theta$ of the sampling distribution that
represents what is known about those parameters after the data $x$ have been observed. We will call
this \emph{probabilistic inference}. A method of inference that is unable to achieve this objective
will be referred to as a procedure for performing \emph{sub-probabilistic inference}.
\end{quote}

The reason that this will be our objective is that those working at the sharp end of scientific
research generally want post-data uncertainty about the unknown true values of parameters to be
expressed in the form of a probability distribution over those parameters. For example, you would
not be much of a weather forecaster if rather than saying what is your probability that there will
be more than one centimetre of rain tomorrow, offered as a substitute some kind of
sub-probabilistic measure of uncertainty in relation to this event, e.g.\ a P value, type I error
rate, confidence interval or likelihood ratio. In this type of scenario, you would expect the
people listening to the forecast to be screaming out `just give us a damn probability!'

Nevertheless, some would disagree that probabilistic inference is a sensible goal due to the fact
that the post-data probabilities that this type of inference requires us to determine will be, in
general, subjective probabilities and it is desirable that, in the interpretation of the results of
a data analysis, any subjectivity is minimised. To counter this criticism, let us point out that
while it is indeed the case that a probability distribution that is placed over a fixed but unknown
quantity will inevitably be in some sense subjective, this does not exclude the possibility that
there could be many who would regard the distribution concerned as being an extremely good and
perhaps almost objective representation of what is known about the quantity of interest.

Also, some would argue that the post-data probability distributions being referred to need to be
validated in terms of their potential sampling performance by comparing the post-data probabilities
that a given parameter of interest will lie in given intervals of the real line with the sampling
probability that these intervals will contain this parameter based on making assumptions about the
underlying sampling distribution.
However, to calculate this sampling probability, an indicator of interval membership will need to
be averaged over a class of cases that, apart from including the case at hand, will also, in
general, include cases that are very different from the case at hand.
For this reason, even under the assumption that this performance checking strategy is applied in
the most sensible manner possible, it will usually be quite easy to justify why a difference may
exist between a given post-data probability of interest and the sampling
probability in question.

Finally, some would claim that a post-data distribution of the parameters $\theta$ can only be
obtained by either explicitly or implicitly inserting a prior distribution for these parameters
into Bayes' theorem, and would then go on to argue that, since the choice of this prior
distribution will in general be difficult to justify, the resulting post-data distribution of the
parameters $\theta$ will not only be subjective but to some extent arbitrary.
To counter this criticism, let us first observe that, in many practical situations, it would be
reasonable to regard an elicited prior distribution of the parameters of interest $\theta$ as being
a very good representation of what was known about these parameters before the data $x$ were
observed, and therefore, in at least these situations, the post-data or posterior distributions of
the parameters $\theta$ that are derived using Bayes' theorem may be regarded as being potentially
quite useful. However, it is perhaps more important to point out that in the theory of
probabilistic inference that will be discussed in the present paper, the use of Bayes' theorem to
derive post-data distributions for parameters of interest will be regarded as being only a special
case of a general approach for obtaining such distributions, i.e.\ a general approach that is
mainly based on non-Bayesian methods.

A difficulty that does arise, though, in developing a theory of probabilistic inference is that
often it will only be possible to determine the post-data probability distributions that are of
interest up to a certain level of precision. Therefore, while in the formal theory of probabilistic
inference that is about to be outlined, it will be assumed that a post-data distribution can be
precisely determined for the parameters $\theta$ of a given sampling distribution, in the practical
application of this theory, we will need to take into account that often there is an upper limit on
how precisely such a distribution can be specified. Nevertheless, it is quite reasonable to treat
this as being a relatively minor issue that, in general, can be adequately resolved by means of a
sensitivity analysis.

By contrast, a formal and comprehensive attempt to express inferences in terms of imprecise
probability, i.e.\ developing a formal theory for determining the lower and upper bounds of the
post-data probability of any given hypothesis about the parameters $\theta$ being true, will be
regarded as being sub-probabilistic inference.
Of course, this means that if this type of sub-probabilistic inference could be considered as being
generally useful, then the assumption we have made concerning the goal of statistical inference
could be brought into question.
Therefore, let us point out that for a formal theory of imprecise probability to be regarded as
being generally useful, a number of important issues have to be addressed. For example, how to give
a real-world meaning to the lower and upper probability bounds concerned, how to determine these
bounds in a non-arbitrary manner, and how to justify that the bounds themselves do not need to be
regarded as being imprecise.
It would appear that these issues represent obstacles that are difficult to overcome.

\vspace{3ex}
\section{Physical and analogical probability}
\label{sec1}

A physical probability of a given event $E$ may be loosely defined as the proportion of outcomes of
an experiment that correspond to the event $E$ occurring when these outcomes have been judged as
being equally likely on the basis of physical symmetry. A precise definition of the concept of
physical probability is given in Bowater~(2022a) in which the meaning of the term `equally likely'
in the loose definition just given is clarified.

Let us now compare the event of randomly drawing out a red ball from an urn that contains a known
number of red balls and a known number of yellow balls with the event of there being more than one
centimetre of rain tomorrow. We may observe that the probability of the first event would usually
be classed as being a physical probability, while this classification would clearly not apply to
the probability of the second event.
Nevertheless, by supposing that the proportion of red balls in the urn can be varied and by making
an analogy between our confidence that the ball drawn out of the urn will be red and our confidence
that there will be more than one centimetre of rain tomorrow, we can both determine our probability
for the latter event and interpret the meaning of this probability.
A more detailed definition of this concept of probability is given in Bowater~(2022a), where a
probability of the type in question is referred to as being an \emph{analogical probability}.

Of course, the analogy between our confidence that the ball drawn out of the urn in the example
just mentioned will be red and our confidence that any given real-world event will occur may, for
example, be strong, not so strong or weak, which leads to the idea that once probabilities of
events or probability distributions of variables have been determined, they can be characterised by
placing them on a scale that reflects the strength of this type of analogy.
Therefore, it is easy to appreciate that, although analogical probabilities are clearly subjective
probabilities that may be considered, in some situations, as being quite arbitrary in nature, they
may nevertheless be regarded, in other situations, as being not that far away from possessing the
objectivity of physical probabilities. A formal characterisation of analogical probabilities in
terms of this concept is given in Bowater~(2022a).

\vspace{3ex}
\section{Overview of the fiducial-Bayes fusion}
\label{sec2}

The theory of inference that will be called the fiducial-Bayes fusion consists of methods of
inference that fall into the following three categories:
\vspace{0.5ex}
\begin{enumerate}
\item Organic fiducial inference used without Bayesian inference
\vspace{-1ex}
\item Organic fiducial inference combined with Bayesian inference
\vspace{-1ex}
\item Standard Bayesian inference
\vspace{0.5ex}
\end{enumerate}
The fiducial-Bayes fusion is therefore a fusion of organic fiducial inference and standard Bayesian
inference that, in some situations, relies on these two methods of inference being used in
combination but, in other situations, respects the capacity of these methods of inference to be
used on their own.

A short summary of organic fiducial inference will be given in the next section, while a longer
discussion of Bayesian inference will be presented in Section~\ref{sec4} that reflects doubts that
may naturally exist concerning the precise role that this latter type of inference should have
within the fiducial-Bayes fusion. Methods of inference that result from combining organic fiducial
inference and Bayesian inference are then discussed in Section~\ref{sec5}.

\vspace{3ex}
\section{Organic fiducial inference}
\label{sec3}

For the purpose of giving a brief overview of organic fiducial inference, let us suppose that we
wish to make inferences about the mean $\mu$ of a normal distribution that has a known variance
$\sigma^2$ on the basis of a sample $x$ of size $n$ drawn from the distribution concerned.
Given that the sample mean $\bar{x}$ is a sufficient statistic for $\mu$, what we call the primary
random variable $\Gamma$ is naturally defined in this case as being:
\pagebreak
\begin{equation}
\label{equ1}
\Gamma = \frac{\bar{x} - \mu}{\sigma/\sqrt{n}}
\vspace{1ex}
\end{equation}
We should clarify that it is convenient to imagine that the value of this variable is generated
directly, e.g.\ by some kind of machine that generates random numbers, and that the sample mean
$\bar{x}$ is determined from this value of $\Gamma$ by simply rearranging equation~(\ref{equ1}).
Notice that the variable $\Gamma$ clearly had a standard normal distribution before the observed
sample mean $\bar{x}$ was generated.

Under the assumption that the strong fiducial argument will be used, the analogy we need to make in
applying organic fiducial inference to this example is with the general argument that if the
outcome of a standard physical experiment, e.g.\ randomly drawing a ball out of an urn of balls or
randomly spinning a wheel, is hidden from us and we receive a specific information packet after the
experiment has taken place, then we may still claim that the probability of any given outcome of
the experiment is the same as it was before the experiment took place if we consider that the
information packet tells us nothing about what the outcome of the experiment might have been.

In applying this analogy to the current example, the outcome of the experiment would be associated
with the realised value of $\Gamma$ and receiving the information packet would be associated with
observing the sample mean $\bar{x}$.
However, if we had held a strong opinion before the data were observed about where on the real line
the true value of $\mu$ was likely to lie, then this analogy is unlikely to be satisfactory as, in
general, observing $\bar{x}$ will change in some way the evaluation we would make of our
uncertainty regarding the true value of $\Gamma$.
Without the support of the analogy that underlies the method of inference that we would be trying
to apply, the use of this method would therefore not be justified.

On the other hand, if we had known nothing or very little about $\mu$ before the data were
observed, then it would not be a surprise if the analogy in question was regarded as being very
acceptable, which as a result, would provide us with an adequate justification for making
inferences about $\mu$ by using the strong fiducial argument.
Just to clarify, using this argument amounts to assuming that $\Gamma$ has the same distribution
that it had before the data $x$ were observed, i.e.\ it follows a standard normal distribution.
This directly implies, by rearranging equation~(\ref{equ1}), that the fiducial or post-data
distribution of $\mu$ is given by the expression:
\pagebreak
\[
\mu\, |\, \sigma^2, x \sim \mbox{N}\hspace{0.05em}(\bar{x},\hspace{0.05em} \sigma^2/n)
\vspace{1ex}
\]

The strong fiducial argument has an important role in statistical inference because:
\par
\begin{enumerate}

\item It can be argued that the most common situation in which we may find ourselves with regard to
our pre-data knowledge about a parameter of interest is one in which the amount of such knowledge
is very limited, and in such a situation, the application of the strong fiducial argument often
works very well.

\item The Bayesian approach to inference often falls down in this type of situation as it is
difficult to express very limited pre-data knowledge about a parameter in the form of a probability
distribution over the parameter.

\end{enumerate}

Furthermore, even in situations where there was a great amount of pre-data knowledge about a
parameter of interest, making post-data inferences about the parameter using organic fiducial
inference will often prove to be more satisfactory than using Bayesian inference.
This is because the theory of organic fiducial inference can incorporate such pre-data knowledge
into the analysis of a data set through what are called global pre-data (GPD) functions and local
pre-data (LPD) functions.
In this way, we can move from using organic fiducial inference under the strong fiducial argument
to using this approach to inference under what are referred to as the moderate and weak fiducial
arguments. We should point out that, whatever amount of pre-data information there is about a
parameter of interest, the manner in which organic fiducial inference incorporates this information
into a data analysis is fundamentally different from the way in which Bayesian inference goes about
achieving this goal through the use of a prior distribution.

The theory of organic fiducial inference was initially developed under the name of `subjective
fiducial inference' in Bowater~(2017, 2018) before being further developed in Bowater~(2019) under
its current name. Certain aspects of this theory were then clarified and modified in
Bowater~(2021). Examples of the application of organic fiducial inference without the use of
Bayesian inference are presented in Bowater~(2018). These examples include cases where the sampling
distribution is a univariate normal, bivariate normal, Pareto, gamma and beta distribution and a
case where the parameter of main interest is a normal mean that depends on a covariate through a
linear regression equation.

\vspace{3ex}
\section{Bayesian inference}
\label{sec4}

\vspace{0.5ex}
\noindent
\textbf{A common viewpoint}

\vspace{1.5ex}
\noindent
Many would consider the Bayesian method to be the most obvious way of obtaining a post-data or
posterior distribution for a parameter of interest. In fact, there are quite a few people who would
go further and claim it is the only way of performing this task.
As a result, they are led to try to find a Bayesian logic behind the assignment of any type of
post-data distribution to the parameters $\theta$ of a given sampling distribution.
The view that this Bayesian logic always will be satisfactory and completely justify the use of the
post-data distribution in question is, however, a flawed view as we will now discuss.

\vspace{4ex}
\noindent
\textbf{Conventional ways of justifying Bayesian inference}

\vspace{1.5ex}
\noindent
Bayes' theorem, which provides the mathematical basis for Bayesian inference, is as follows:
\vspace{1.5ex}
\[
P(A\,|\,B) = \frac{P(B\,|\,A)P(A)}{P(B)}
\vspace{2ex}
\]
where $A$ and $B$ are two events. Observe that if the events $A$ and $B$ are generated from the
joint distribution of these two events through the use of a standard physical experiment of the
type discussed in Sections~\ref{sec1} and~\ref{sec3}, then the application of Bayes' theorem is
almost without controversy and feels completely intuitive.

However, when Bayes' theorem is applied in practice to the problem of making post-data inferences
about the parameters $\theta$ of a sampling distribution, the event $A$ is not, in general,
randomly generated by a well-understood experiment but instead corresponds to a fixed but unknown
state of the world, i.e.\ the fixed but unknown parameters of interest.
Therefore, in these circumstances, rather than being, in effect, a summary of the technical
specifications of a physical experiment, the prior distribution $P(A)$ is merely an expression of
our pre-data knowledge about the parameters of interest $\theta$.
This naturally leads to the question of what justification we have for applying Bayes' theorem in
the latter case rather than the former case. Despite the popularity of Bayesian inference, it is,
in fact, not that easy to answer this question.

From the earliest works in the development of Bayesian inference, e.g.\ Ramsey~(1926),
de Finetti~(1937) and Savage~(1954), it has, of course, been accepted that a substantial
justification is required for the use of Bayes' theorem in the case where the prior distribution
merely represents pre-data knowledge about a fixed but unknown quantity.
Efforts in this direction have, quite naturally, focused on constructing systems of human behaviour
based on specific assumptions or axioms with the aim that, if an individual adheres to these
assumptions or axioms, then the way that they draw inferences from any given data set must be in
accordance with Bayes' theorem (see in addition to the works just cited, Fishburn~1986, Bernardo
and Smith~1994 and Jaynes~2003).
However, if we consider the acceptability of the axioms on which these systems are based in all
relevant circumstances, not just in the most obvious circumstances, then in summary, it can be
strongly argued that the level of intuition required to accept these axioms is not genuinely less
than the level of intuition required to directly accept that Bayes' theorem should always be
followed when drawing inferences from data.
Therefore, it is reasonable to doubt that the systems of human behaviour under discussion achieve
in any way their intended goal of justifying the general applicability of Bayes' theorem.

\vspace{4ex}
\noindent
\textbf{Justifying Bayesian inference through analogy}

\vspace{1.5ex}
\noindent
In any given situation where we have the option of applying Bayes' theorem to update a subjective
prior distribution for a set of fixed but unknown parameters $\theta$ to a posterior distribution
for these parameters, we may, instead of looking for a general way of justifying the application of
this theorem, choose to try to justify the use of this theorem only in this particular situation by
making an analogy between the variables of interest, i.e.\ the parameters $\theta$ and the data
$x$, and the outcomes of the type of standard physical experiment to which Bayes' theorem is most
naturally applied. Let us from now on refer to this type of analogy as Bayes' analogy. The use of
this term is indeed very appropriate as Thomas Bayes himself used such an analogy to justify the
use of Bayes' theorem to make post-data inferences about a binomial proportion in his famous paper
that originally presented this theorem, i.e.\ Bayes~(1763), with his choice of standard physical
experiment being the random throwing of balls onto a square table.

Using Bayes' theorem to make post-data inferences about fixed but unknown parameters of interest on
a case-by-case basis through Bayes' analogy would appear, at least at first sight, to be an
intuitive and sensible way of making use of this theorem. Nevertheless, adopting this strategy
rather than endorsing the use of Bayes' theorem as being the only acceptable or rational way of
making inferences of the type in question does have some quite fundamental consequences.

For example, if we are presented with a post-data distribution for the parameters of interest
$\theta$ that has been derived using a non-Bayesian method, e.g.\ using organic fiducial inference,
are we entitled to use Bayes' theorem in reverse to discover what was the prior distribution for
these parameters?
Here, we would of course be entitled to use Bayes' theorem in reverse if we accept the use of
Bayes' analogy in the given context. However, if the post-data distribution of the parameters in
question was not derived using Bayes' theorem because the use of Bayes' analogy was not accepted in
that context, i.e.\ in the forward direction, why should we be willing to regard Bayes' analogy as
being acceptable in going in the reverse direction?

Also, if the use of Bayes' theorem in any given context is optional and depends on the
acceptability of making Bayes' analogy in that context, then we may quite reasonably wish to know
what are the criteria we may use for evaluating whether or not making this analogy is acceptable or
not. To explore this issue, let us make the seemingly quite natural assumption that, given a prior
distribution $p(\theta)$ for the parameters $\theta$ of a given sampling distribution, Bayes'
analogy should be considered a good analogy to make if a good analogy can be made between our
pre-data uncertainty about the parameters $\theta$ in the real-world context of interest and what
our uncertainty about these parameters would be if they were about to be generated from the prior
distribution $p(\theta)$ using a standard physical experiment.
One way of evaluating whether this second type of analogy is a good analogy to make is to evaluate
the similarity there is between the confidence we would have had, before the data were observed,
that the parameters $\theta$ would lie in given regions of the parameter space and our confidence
that, when generated using the standard physical experiment in question, these parameters would lie
in each of the regions concerned.
If there is a high degree of similarity between our confidence in events of these two types
occurring, then this is a loose way of saying that the prior distribution $p(\theta)$ would be
regarded as being \emph{externally strong} according to the terminology and definitions given in
Bowater~(2022a) that relate to the concept of analogical probability discussed in
Section~\ref{sec1}.

Nevertheless, just because in any given context, Bayes' analogy is evaluated as being a good
analogy to make according to this particular criterion does not mean that we should regard the
conclusions that result from substituting the prior distribution $p(\theta)$ concerned into Bayes'
theorem as being trustworthy.
This is simply due to the fact that, in the scenario of interest, the prior distribution
$p(\theta)$ will always be a subjective probability distribution, and therefore, will always fall
short of being the kind of distribution it would be if the true value of the parameter $\theta$ was
determined by setting it equal to the outcome of a standard physical experiment.
To emphasize this point, let us now consider two counter-examples to the idea that, in any given
context, the criterion just discussed for evaluating the quality of Bayes' analogy is always going
to be adequate.

\vspace{4ex}
\noindent
\textbf{Two counter-examples}

\vspace{1.5ex}
\noindent
The first example of the type just mentioned that we will highlight is known as `Lindley's
paradox', which was originally discussed in Jeffreys~(1939) before being first called a paradox in
Lindley~(1957).
This example shows that if the sampling distribution is, say, a normal distribution centred at a
given parameter $\nu$ then, for large sample sizes, the standard two-sided P value for the null
hypothesis that the parameter $\nu$ equals a given value $\nu_0$ can be very small, and yet, the
posterior probability of this hypothesis being true can be close to one even when only a small
prior probability is placed on this hypothesis and the prior distribution over other values of the
parameter $\nu$ is concentrated around the value $\nu_0$.
Although the role of P values in statistical inference can be questioned, the way in which, in this
example, pre-data knowledge about the parameter $\nu$ is updated to post-data knowledge about this
parameter by using Bayes' theorem would be clearly regarded, in most practical situations, as being
completely inappropriate, with an important exception being, of course, the case where $\nu$ was in
fact generated from the prior distribution concerned by using a standard physical experiment.

While the example just given relates to the use of Bayes' theorem in quite a specific context,
i.e.\ a null hypothesis testing problem, another example highlighted by Lindley and published a
year later in Lindley~(1958) relates to a more general context.
In this example, what will be called, for the moment, method of inference A is used to form a
distribution for a parameter of interest $\tau$ that represents our knowledge about this parameter
after having observed a data set $x$, and also, the same method A is used to form a distribution
for $\tau$ that represents our knowledge about this parameter after having observed both the data
set $x$ and an additional data set $y$.
It is then shown that, in quite a general class of cases, this latter post-data distribution of
$\tau$ is not equal to the posterior distribution of $\tau$ that is formed by treating the former
post-data distribution of $\tau$ as a prior distribution of $\tau$ in a Bayesian analysis of the
data set $y$.
Of course, this example can be used, and has been used, to argue that the method of inference A
should be regarded as being flawed.
However, if we imagine that there was no particular reason to question the adequacy of the method
of inference A, then since, as already discussed, the justifications for using Bayes' theorem in
the general context of interest are rather weak, this example should clearly be regarded as being a
counter-example to the use of Bayes' theorem in this context.

The method of inference A just referred to is, in fact, a simple special case of the general method
of organic fiducial inference that was summarised in Section~\ref{sec3}.
Furthermore, since it can be strongly argued that the way organic fiducial inference is used in the
example under discussion enables us to obtain a post-data distribution of the parameter $\tau$ that
is a very good representation of all that can be known about this parameter after having observed
the data being analysed, it can be strongly argued that we have just identified another
counter-example to the use of Bayes' theorem in the case where the prior distribution is considered
to be a very good representation of our pre-data knowledge about the parameter of interest.

\vspace{4ex}
\noindent
\textbf{Conclusions about the role of Bayesian inference}

\vspace{1.5ex}
\noindent
In summary, it would appear difficult to establish the precise conditions under which the
application of Bayes' theorem should be considered appropriate in the case where the prior
distribution merely represents pre-data knowledge about fixed but unknown parameters of interest,
and therefore, the decision to use Bayes' theorem in this case would always seem to require, to
some degree, `a leap of faith'.
Nevertheless, minimum requirements, i.e.\ necessary conditions, for a successful application of
Bayes' theorem in the case in question would appear to be as follows:

\vspace{2ex}
\noindent
1) We are able to place a probability distribution over the parameters of interest $\theta$ that
adequately represents what we knew about these parameters before the data were observed.

\vspace{2ex}
\noindent
2) We can assure ourselves that the counter-examples that exist to the use of Bayes' theorem in the
case where the prior distribution is of the general nature being considered do not either fully or
partially apply to the particular case being studied.

\vspace{2ex}
On the other hand, it would appear undeniable that, within the general class of scenarios where the
prior distribution merely represents pre-data knowledge about fixed but unknown parameters of
interest $\theta$, cases exist where choosing to make inferences about the parameters $\theta$ by
using Bayes' theorem may be intuitively regarded as being quite a sensible way to proceed.
Therefore, we may conclude that the precise contribution Bayesian inference is capable of making in
addressing real-world problems of inference through its role in the general theory of inference
being discussed in the present paper, i.e.\ the fiducial-Bayes fusion, should be considered as
being, to some extent, an open issue.

\vspace{3ex}
\section{Organic fiducial inference combined with Bayesian inference}
\label{sec5}

\vspace{0.5ex}
\noindent
\textbf{Single parameter problems}

\vspace{1.5ex}
\noindent
In the case where the sampling distribution depends on only one unknown parameter, the theory of
organic fiducial inference is naturally extended in Bowater~(2021) to the problem of making
inferences about the parameters of discrete sampling distributions by using Bayesian inference to
make inferences about the parameter of interest within given closed intervals of the parameter
space. However, since these intervals are usually quite narrow, this particular way of
incorporating Bayesian inference into the theory of organic fiducial inference has generally quite
a small impact on the overall post-data conclusions that are drawn about the parameter of interest.
Examples of the application of the method of inference in question in cases where the sampling
distribution is assumed to be a binomial, Poisson and multinomial distribution are outlined in
Bowater~(2021).

\pagebreak
\noindent
\textbf{Extending the fiducial-Bayes fusion to multiparameter problems}

\vspace{1.5ex}
\noindent
A first step of a general strategy for extending the whole class of methods of inference that
belong to the fiducial-Bayes fusion from single parameter to multiparameter problems is to
determine the set of full conditional post-data densities for the parameters
$\theta = \{\theta_1, \theta_2, \ldots, \theta_k\}$ of the sampling distribution of interest, i.e.\
the set of full conditional densities:
\vspace{0.25ex}
\begin{equation}
\label{equ2}
p(\theta_j\,|\,\theta_{-j},x)\ \ \ \mbox{for $j=1,2,\ldots,k$}
\vspace{0.75ex}
\end{equation}
(where $\theta_{-j}$ denotes all the parameters in $\theta$ except for $\theta_j$) with it being
understood that any given density function in this set is determined using whichever method of
inference belonging to the fiducial-Bayes fusion that is considered to be the most appropriate for
that particular task.
If these full conditional densities are compatible, then under a mild condition, they determine a
unique joint post-data density for the parameters $\theta$, and indeed, this joint density function
can sometimes be derived using analytical methods.

On the other hand, if the full conditional densities in equation~(\ref{equ2}) are not consistent
with any joint distribution of the parameters $\theta$, in other words, they are incompatible, then
very often they will be approximately compatible, i.e.\ a joint density function of the parameters
$\theta$ can be found that has full conditional densities that closely approximate those given in
equation~(\ref{equ2}).
A way of determining the joint density of the parameters $\theta$ that best fits this criterion is
by using a Gibbs sampler based on the full conditional densities in equation~(\ref{equ2}) with an
appropriately chosen scanning order of the parameters $\theta$. (Note that the limiting
distribution of a Gibbs sampler is affected by the scanning order of the parameters if the full
conditional densities on which it is based are not compatible.)

A detailed account of this latter approach is presented in both Bowater~(2018) and Bowater~(2020).
Clearly, if at least one of the full conditional post-data densities in equation~(\ref{equ2}) is
determined by using organic fiducial inference and at least of these full conditional densities is
determined by using Bayesian inference, then the method of inference that results from applying the
general strategy just outlined can naturally be regarded as being a combination of organic fiducial
inference and Bayesian inference.
Examples of this particular way of combining organic fiducial inference and Bayesian inference can
be found in Bowater~(2020). These examples include \pagebreak cases where the sampling distribution
is a univariate normal and trinomial distribution. Moreover, it should be pointed out that another
example of the type in question in which the sampling distribution is a bivariate normal
distribution is laid out in the Appendix of the present paper.

\vspace{4ex}
\noindent
\textbf{Testing sharp or almost sharp hypotheses about the parameters of interest}

\vspace{1.5ex}
\noindent
Let us suppose that there was a substantial degree of pre-data belief in the hypothesis that the
values of the parameters $\theta$ would lie in a given small region $R$ of the parameter space, and
that, in order to represent this pre-data knowledge about the parameters $\theta$ adequately, a
given pre-data probability has been assigned to the hypothesis in question.
Also, we will assume that both in the case where the parameters $\theta$ are conditioned to lie in
the region $R$ and in the case where these parameters are conditioned not to lie in this region,
our pre-data knowledge about the parameters $\theta$ can be adequately expressed in the way we are
allowed to express this knowledge in the core theory of organic fiducial inference that is outlined
in Bowater~(2019, 2021).

Given that our pre-data knowledge about the parameters $\theta$ has been expressed in the two
distinct ways that have just been described, we may form a joint probability distribution of the
parameters $\theta$ that adequately represents our knowledge about these parameters after having
observed a data set $x$ by combining organic fiducial inference with Bayesian inference in the way
that is carefully detailed in Bowater~(2022b).
The method of inference being referred to is, therefore, another example of a method of inference
that would be classified as belonging to item 2 in the list of approaches to statistical inference
given in Section~\ref{sec2} that constitute what has been termed the fiducial-Bayes fusion.
Examples of the application of the method of inference in question to the testing of sharp or
almost sharp hypotheses in cases where the parameter of main interest is a normal mean, binomial
proportion and relative risk can be found in Bowater~(2022b).

\vspace{3ex}
\section{Closing remark}

The theory of inference that has been called the fiducial-Bayes fusion can be viewed as lying
within a field of study that could be given the name `The general updating of probabilistic
knowledge' (TGUPK).
The root of this name can be thought of as being found in the name `The Bayesian updating of
probabilistic knowledge', which can be naturally assumed to imply nothing more or less than
Bayesian inference. Therefore, the field of TGUPK can be regarded as containing both Bayesian and
non-Bayesian methods for updating pre-data probability distributions of quantities of interest to
post-data distributions for these quantities.
It is reasonable to hope that methods of inference belonging to the fiducial-Bayes fusion will be
regarded as being amongst the most adequately justified and useful methods of inference that can be
found in the field of TGUPK.

\vspace{3ex}
\noindent
\textbf{Note:} A fairly long appendix to the present paper, which contains a substantive
ex\-am\-ple of the application of the theory of the fiducial-Bayes fusion, can be found after the
bibliography.

\vspace{5.5ex}
\pdfbookmark[0]{Bibliography}{toc1}
\noindent
\textbf{Bibliography}

\begin{description}

\setlength{\itemsep}{1ex}

\vspace{0.5ex}
\item[] Bayes, T. (1763).\ An essay towards solving a problem in the doctrine of chances.\
\emph{Philosophical Transactions of the Royal Society}, \textbf{53}, 370--418.

\item[] Bernardo, J. M. and  Smith, A. F. M. (1994).\ \emph{Bayesian Theory}, Wiley, New York.

\item[] Bowater, R. J. (2017).\ A defence of subjective fiducial inference.\ \emph{AStA Advances in
Statistical Analysis}, \textbf{101}, 177--197.

\item[] Bowater, R. J. (2018).\ Multivariate subjective fiducial inference.\ \emph{arXiv.org
(Cornell University), Statistics}, arXiv:1804.09804.

\item[] Bowater, R. J. (2019).\ Organic fiducial inference.\ \emph{arXiv.org (Cornell University),
Sta\-tis\-tics}, arXiv:1901.08589.

\item[] Bowater, R. J. (2020).\ Integrated organic inference (IOI):\ a reconciliation of
statistical paradigms.\ \emph{arXiv.org (Cornell University), Statistics}, arXiv:2002.07966.

\item[] Bowater, R. J. (2021).\ A revision to the theory of organic fiducial inference.\
\emph{arXiv.org (Cornell University), Statistics}, arXiv:2111.09279.

\item[] Bowater, R. J. (2022a).\ Physical, subjective and analogical probability.\ \emph{arXiv.org
(Cornell University), Statistics}, arXiv:2204.10159.

\item[] Bowater, R. J. (2022b).\ Sharp hypotheses and organic fiducial inference.\ \emph{arXiv.org
\\ (Cornell University), Statistics}, arXiv:2207.08882.

\item[] Brooks, S. P. and Roberts, G. O. (1998).\ Convergence assessment techniques for Markov
chain Monte Carlo.\ \emph{Statistics and Computing}, \textbf{8}, 319--335.

\item[] de Finetti, B. (1937).\ La pr\'evision:\ ses lois logiques, ses sources subjectives.\
\emph{Annales de l'Institut Henri Poincar\'e}, \textbf{7}, 1--68.\ Translated into English in 1964
as `Foresight:\ its logical laws, its subjective sources' in \emph{Studies in Subjective
Probability}, Eds.\ H. E. Kyburg and H. E. Smokler, Wiley, New York, pp.\ 93--158.

\item[] Efron, B. (1993).\ Bayes and likelihood calculations from confidence intervals.\
\emph{Bio\-met\-rika}, \textbf{80}, 3--26.

\item[] Fishburn, P. C. (1986).\ The axioms of subjective probability (with discussion).\
\emph{Sta\-tis\-ti\-cal Science}, \textbf{1}, 335--358.

\item[] Gelman, A. and Rubin, D. B. (1992).\ Inference from iterative simulation using multiple
sequences.\ \emph{Statistical Science}, \textbf{7}, 457--472.

\item[] Jaynes, E. T. (2003).\ \emph{Probability Theory:\ The Logic of Science}, Cambridge
University Press, Cambridge.

\item[] Jeffreys, H. (1939).\ \emph{Theory of Probability}, 1st edition, Clarendon Press, Oxford.

\item[] Lindley, D. V. (1957).\ A statistical paradox.\ \emph{Biometrika}, \textbf{44}, 187--192.

\item[] Lindley, D. V. (1958).\ Fiducial distributions and Bayes' theorem.\ \emph{Journal of the
Royal Statistical Society, Series B}, \textbf{20}, 102--107.

\item[] Ramsey, F. P. (1926).\ \emph{Truth and Probability}.\ Originally an unpublished
manuscript.\ Presented in 1964 in \emph{Studies in Subjective Probability}, Eds.\ H. E. Kyburg and
H. E. Smokler, Wiley, New York, pp.\ 61--92.

\item[] Savage, L. J. (1954).\ \emph{The Foundations of Statistics}, Wiley, New York.

\end{description}

\pagebreak
\pdfbookmark[0]{Appendix}{toc2}
\noindent
\textbf{Appendix}

\vspace{2ex}
\noindent
\textbf{Supplementary example}

\vspace{1.5ex}
\noindent
To give an additional example to the examples that were referenced in the present paper of how the
theory of the fiducial-Bayes fusion can be applied, let us consider the problem of making
inferences about the five parameters of a bivariate normal distribution on the basis of a random
sample drawn from this type of distribution.
To clarify, we will be interested in making inferences about the means $\mu_x$ and $\mu_y$ and the
variances $\sigma^2_x$ and $\sigma^2_y$ of the two random variables concerned $X$ and $Y$,
respectively, and the correlation $\rho$ of $X$ and $Y$ on the basis of an observed
sample\hspace{0.03em} $\mathtt{z}=\{(x_i,y_i) : i=1,2,\ldots,n\}$, where $x_i$ and $y_i$ are the
$i$th realisations of $X$ and $Y$, respectively.

In Bowater~(2018), as a way of addressing this problem, full conditional fiducial densities were
derived either exactly or approximately for each of the parameters $\mu_x$, $\mu_y$, $\sigma^2_x$,
$\sigma^2_y$ and $\rho$ by using appropriately chosen fiducial statistics under the strong fiducial
argument, and then it was illustrated how, on the basis of these conditional densities, what can be
regarded as being a suitable joint fiducial density of these parameters can be obtained by using
the Gibbs sampler within the type of framework summarised in Section~\ref{sec5} of the current
paper. However, while we will again attempt to apply this general framework to the problem of
interest, the specific method just referred to is not going to be directly applicable to the case
that will be presently considered.
This is because, although it will be assumed that nothing or very little was known about the
correlation coefficient $\rho$ before the data were observed, by contrast it is going to be assumed
that there was a substantial amount of pre-data knowledge about the means $\mu_x$ and $\mu_y$ and
the variances $\sigma^2_x$ and $\sigma^2_y$.
To begin with though, let us clarify how, by using the theory of organic fiducial inference, the
full conditional post-data density of $\rho$ will be constructed.

In this regard, observe that, if all the parameters except $\rho$ are known, there exists no
sufficient set of univariate statistics for $\rho$ that contains only one statistic that is not an
ancillary statistic, and therefore the simple definition of a fiducial statistic given in
Bowater~(2019) can not be applied in this case.
However, according to the more general definition of a fiducial statistic given in Bowater~(2021),
it would seem reasonable to assume that the fiducial statistic in this case is the maximum
likelihood estimator of $\rho$ given that all other parameters are known.
Indeed, this was the fiducial statistic that was used in a similar scenario to the one being
discussed in the example in Bowater~(2018) that was just mentioned.

It can be shown that the maximum likelihood estimator of $\rho$ in question is the value
$\bm\hat{\rho}$ that solves the following cubic equation:
\vspace{1ex}
\[
-n\bm\hat{\rho}^{\hspace{0.05em}3} + \left( \frac{\sum_{i=1}^n x'_i y'_i}{\sigma_x \sigma_y}
\right)\hspace{-0.1em} \bm\hat{\rho}^{\hspace{0.05em}2} + \left( n -
\frac{\sum_{i=1}^n (x'_i)^2}{\sigma_x^2} - \frac{\sum_{i=1}^n (y'_i)^2}{\sigma_y^2} \right)
\hspace{-0.1em} \bm\hat{\rho}\hspace{0.05em} + \hspace{0.03em}
\frac{\sum_{i=1}^n x'_i y'_i}{\sigma_x \sigma_y} = 0
\vspace{1ex}
\]
Of course, it is well known that a maximum likelihood estimator of a parameter is usually
asymptotically normally distributed with mean equal to the true value of the parameter, and
variance equal to the inverse of the Fisher information with respect to that parameter.
(To clarify, this is the Fisher information obtained via differentiating the logarithm of the
likelihood function with respect to the parameter concerned.) For this reason, if $n$ is large, the
sampling distribution of the maximum likelihood estimator $\bm\hat{\rho}$ just defined can be
approximately expressed as follows:
\vspace{-0.25ex}
\[
\bm\hat{\rho} \sim
\mbox{N}\hspace{0.05em}(\hspace{0.05em}\rho,\hspace{0.05em} 1/ \mathcal{I}(\rho)\hspace{0.05em})
\vspace{-0.25ex}
\]
where the value $\mathcal{I}(\rho)$ is the Fisher information of the likelihood function in this
example with respect to $\rho$ assuming all other parameters are known, which is in fact a value
that is given by the expression:
\vspace{2ex}
\[
\mathcal{I}(\rho) = \frac{n(1+\rho^{\hspace{0.03em}2})}{(1-\rho^{\hspace{0.04em}2})^2}
\vspace{2ex}
\]

This is the way in which the sampling distribution of the estimator $\bm\hat{\rho}$ was
approximated in the context of current interest in Bowater~(2018). However, instead of using this
approximation, let us assume that it is a transformation of $\bm\hat{\rho}$, namely the function
$\tanh^{-1} (\bm\hat{\rho})$, rather than simply the estimator $\bm\hat{\rho}$ that is normally
distributed.
The reason for doing this is that it can be shown that, under this alternative assumption, a
generally better approximation to the sampling density of $\bm\hat{\rho}$ can be obtained than
under the original assumption, except, that is, when $\rho$ is close to zero.
As a result, using the conventional method being discussed, it will be assumed that the density
function of $\tanh^{-1}(\bm\hat{\rho})$ is directly specified, and the density function of
$\bm\hat{\rho}$ is therefore indirectly specified, by the expression:
\vspace{1.5ex}
\[
\tanh^{-1}(\bm\hat{\rho}) \sim \mbox{N}\hspace{0.05em} (\hspace{0.05em}\tanh^{-1}(\rho),
\hspace{0.05em}1/ \mathcal{I}(\tanh^{-1}\rho)\hspace{0.05em})
\vspace{1.25ex}
\]
where the value $\mathcal{I}(\tanh^{-1} \rho)$ is the Fisher information with respect to the
quantity $\tanh^{-1}(\rho)$ assuming all parameters except $\rho$ are known, which is in fact a
value that can be expressed as follows:
\vspace{-0.5ex}
\[
\mathcal{I}(\tanh^{-1} \rho) = n(1+\rho^2)
\]

Allowing the statistic $\tanh^{-1}(\bm\hat{\rho}$) to take the role of the fiducial statistic, and
using the approximation to the sampling density of this statistic just given, we can therefore
approximate, in the present case, the general equation used to determine any given fiducial
statistic as part of the data generating algorithm detailed in Bowater~(2019, 2021) in the
following way:
\vspace{0.5ex}
\begin{equation}
\label{equ3}
\tanh^{-1}(\bm\hat{\rho}) = \varphi(\Gamma,\rho) = \tanh^{-1}(\rho) +
\frac{\Gamma}{\sqrt{n(1+\rho^{\hspace{0.04em}2})}}
\vspace{1.25ex}
\end{equation}
where the primary random variable $\Gamma$ has a standard normal distribution.
Although it can be shown that, given a value for $\bm\hat{\rho}$, this equation does not generally
define a bijective mapping between all possible values of $\Gamma$ and all possible values of
$\rho$, it is the case, on the other hand, that if $\Gamma$ is generated from a standard normal
density function truncated to lie in a given interval $[-\alpha,\alpha]$ where $\alpha>0$, then a
bijective mapping of this nature does exist for very large values of $\alpha$ under the restriction
that $n$ is not too small and $\bm\hat{\rho}$ is not very close to $-1$ or $1$.
For example, if $n=100$ and $|\bm\hat{\rho}| < 0.999$, then this type of bijective mapping will
exist not only for small values of $\alpha$, but also if $\alpha$ is chosen to be as high as 36,
and will exist for substantially larger values of $\alpha$ as $|\bm\hat{\rho}|$ becomes smaller.

We will therefore make use of equation~(\ref{equ3}) under the assumption that the primary random
variable $\Gamma$ follows the truncated normal density function just mentioned with $\alpha$ chosen
to be equal to or not far below the largest possible value of $\alpha$ that is consistent with
equation~(\ref{equ3}) satisfying Condition~1 of Bowater~(2019). Clearly, if this condition can be
approximately satisfied in this way then, also in an approximate sense, we may apply Principle~1
for deriving the fiducial density of a general parameter that was outlined in Section~3.4 of this
earlier paper.
Furthermore, since the fiducial density of $\rho$ given all other parameters are known needs to be
derived under the \pagebreak assumption that, given the values of the means $\mu_x$ and $\mu_y$ and
the variances $\sigma_x^2$ and $\sigma_y^2$, there would have been no or very little pre-data
knowledge about the correlation coefficient $\rho$, it will be quite naturally assumed that what
was defined, in Bowater~(2019, 2021), as being a global pre-data (GPD) function is specified in the
present case as being a function of $\rho$ that is equal to a positive constant over the interval
$[-1,1]$ and equal to zero otherwise.
Under the assumptions that have just been made, applying Principle~1 of Bowater~(2019), which was
just referred to, leads to an approximation to the full conditional fiducial density of $\rho$ that
is given by:
\begin{equation}
\label{equ7}
f(\rho\,|\,\mu_x,\mu_y,\sigma_x^2,\sigma_y^2,\mathtt{z}) = \left\{
\begin{array}{ll}
{\displaystyle \psi_{\alpha}(\gamma)  \left| \frac{d\gamma}{d\rho} \right|}\,
& \mbox{if $\rho \in [\hspace{0.05em}\rho_0, \rho_1]$}\\[2.5ex]
0 & \mbox{otherwise}
\end{array}
\right.
\vspace{1.5ex}
\end{equation}
where $\gamma$ is the value of $\Gamma$ that solves equation~(\ref{equ3}) for the given value of
$\rho$, i.e.\
\vspace{0.25ex}
\[
\gamma = (\tanh^{-1}(\bm\hat{\rho}) - \tanh^{-1}(\rho)) \sqrt{n(1+\rho^{\hspace{0.04em}2})}
\]
while $\psi_{\alpha}(\gamma)$ is the standard normal density function truncated to lie in the
interval $[-\alpha,\alpha]$ evaluated at $\gamma$, and finally $[\hspace{0.05em}\rho_0, \rho_1]$ is
the interval of values of $\rho$ that, according to equation~(\ref{equ3}), correspond to $\gamma$
lying in the interval $[-\alpha,\alpha]$.

With regard to what was known about the means $\mu_x$ and $\mu_y$ and the variances $\sigma_x^2$
and $\sigma_y^2$ before the data set $\mathtt{z}$ was observed, we will assume that it is possible
to adequately represent such pre-data knowledge by placing a probability density function over each
of these parameters conditional on all parameters except the parameter itself being known, i.e.\
the density functions $p(\mu_x\,|\,\mu_y,\sigma_x^2,\sigma_y^2,\rho)$,
$p(\mu_y\,|\,\mu_x,\sigma_x^2,\sigma_y^2,\rho)$, $p(\sigma_x^2\,|\,\mu_x,\mu_y,\sigma_y^2,\rho)$
and $p(\sigma_y^2\,|\,\mu_x,\mu_y,\sigma_x^2,\rho)$, respectively.
To give an example, let the full conditional densities of $\mu_x$ and $\mu_y$ in question be
defined by:
\begin{equation}
\label{equ4}
\mu_x \sim \mbox{N}\hspace{0.05em}(\hspace{0.05em}\mu'_x,\hspace{0.05em} \sigma_x^2/n'_x)\ \
\mbox{and}\ \
\mu_y \sim \mbox{N}\hspace{0.05em}(\hspace{0.05em}\mu'_y,\hspace{0.05em} \sigma_y^2/n'_y)
\vspace{0.25ex}
\end{equation}
respectively, where $\mu'_x \in \mathbb{R}$, $n'_x \geq 2$, $\mu'_y \in \mathbb{R}$ and
$n'_y \geq 2$ are given constants, and let the full conditional densities of $\sigma_x^2$ and
$\sigma_y^2$ in question be defined by:
\begin{equation}
\label{equ5}
\sigma_x^2 \sim \mbox{Inv-Gamma}\, (n'_x/2,\hspace{0.05em} \beta_x)\ \ \mbox{and}\ \
\sigma_y^2 \sim \mbox{Inv-Gamma}\, (n'_y/2,\hspace{0.05em} \beta_y)
\pagebreak
\end{equation}
respectively, where
\[
\beta_x = (1/2)(n'_x-1)(\sigma'_x)^2 + (n'_x/2)(\mu_x-\mu'_x)^2\ \ \mbox{and}\ \
\beta_y = (1/2)(n'_y-1)(\sigma'_y)^2 + (n'_y/2)(\mu_y-\mu'_y)^2
\vspace{0.25ex}
\]
in which $\sigma'_x>0$ and $\sigma'_y>0$ are given constants, i.e.\ the densities of $\sigma_x^2$
and $\sigma_y^2$ being referred to are inverse gamma density functions with shape parameters equal
to $n'_x/2$ and $n'_y/2$ and scale parameters equal to $\beta_x$ and $\beta_y$, respectively.
These choices for what, under the Bayesian paradigm, would be referred to as the full conditional
prior densities of $\mu_x$, $\mu_y$, $\sigma_x^2$ and $\sigma_y^2$ have practical relevance since
they are consistent with the joint prior density of $\mu_x$ and $\sigma_x^2$ and the joint prior
density of $\mu_y$ and $\sigma_y^2$ being equal, respectively, to the joint fiducial density of
$\mu_x$ and $\sigma_x^2$ and the joint fiducial density of $\mu_y$ and $\sigma_y^2$ that would be
formed, according to the generic definition of the two types of joint fiducial density being
referred to given in Section~5 of Bowater~(2019), on the basis of a preliminary sample of $n'_x$
realisations of the variable $X$ and an independent preliminary sample of $n'_y$ realisations of
the variable $Y$.
In particular, the conditional prior densities of $\mu_x$, $\mu_y$, $\sigma_x^2$ and $\sigma_y^2$
defined in equations~(\ref{equ4}) and~(\ref{equ5}) are consistent with the former preliminary
sample having a mean of $\mu'_x$ and a standard deviation of $\sigma'_x$ and the latter preliminary
sample having a mean of $\mu'_y$ and a standard deviation of $\sigma'_y$.

It is clear from the definition of the density function for a bivariate normal distribution that
the likelihood function of the parameters $\mu_x$, $\mu_y$, $\sigma_x^2$ and $\sigma_y^2$ for the
example of current interest in the case where the correlation $\rho$ is known can be expressed as
follows:
\vspace{1.25ex}
\begin{eqnarray}
\label{equ6}
L(\mu_x,\mu_y,\sigma_x^2,\sigma_y^2\,|\,\rho,\mathtt{z}) & = & (1/\sigma_x\sigma_y)^n \exp\left(
\frac{-1}{2(1-\rho^{\hspace{0.04em}2})}\left( \frac{\sum (x'_i)^2}{\sigma_x^2} \right) \right.
\nonumber\\[1.5ex]
&& + \left. \frac{\rho}{1-\rho^{\hspace{0.04em}2}} \left(\frac{\sum x'_i y'_i}{\sigma_x \sigma_y}
\right) - \frac{1}{2(1-\rho^{\hspace{0.04em}2})}\left( \frac{\sum (y'_i)^2}{\sigma_y^2} \right)
\right)
\end{eqnarray}
\par \vspace{1.5ex} \noindent
where $x'_{i} = x_{i} - \mu_x$ and $y'_{i} = y_{i} - \mu_y$.
Therefore, if the density functions of $\mu_x$, $\mu_y$, $\sigma_x^2$ and $\sigma_y^2$ specified in
equations~(\ref{equ4}) and~(\ref{equ5}) are treated as full conditional prior densities, it can
easily be seen how, by using Bayes' theorem on the basis of these prior densities and the
likelihood function in equation~(\ref{equ6}), the full conditional posterior densities of $\mu_x$,
$\mu_y$, $\sigma_x^2$ and $\sigma_y^2$ can be numerically computed, i.e.\ the posterior densities
$p(\mu_x\,|\,\mu_y,\sigma_x^2,\sigma_y^2,\rho,\mathtt{z})$,
$p(\mu_y\,|\,\mu_x,\sigma_x^2,\sigma_y^2,\rho,\mathtt{z})$,
$p(\sigma_x^2\,|\,\mu_x,\mu_y,\sigma_y^2,\rho,\mathtt{z})$ and \pagebreak
$p(\sigma_y^2\,|\,\mu_x,\mu_y,\sigma_x^2,\rho,\mathtt{z})$.
These full conditional posterior densities together with the full conditional fiducial density of
$\rho$ defined by equation~(\ref{equ7}) form a complete set of full conditional post-data densities
for the example being studied, i.e.\ a set of conditional densities of the type on which the
strategy for constructing a joint post-data density of a general set of parameters $\theta$
outlined in Section~\ref{sec5} of the present paper is based.

To develop the analysis of this example further, Figure~1 shows some results from running a Gibbs
sampler with a uniform random scanning order of the parameters $\mu_x$, $\mu_y$, $\sigma_x^2$,
$\sigma_y^2$ and $\rho$ on the basis of the full conditional post-data densities of these
parameters that have just been detailed.
To clarify, each transition of this Gibbs sampling algorithm was the result of generating a value
from one of the full conditional densities under discussion that was chosen at random, with the
same probability of $1/5$ being given to any one of these densities being selected, and then
treating the generated value as the updated value of the parameter concerned.
The histograms in Figures~1(a) to~1(e) represent, in particular, the distributions of the values of
$\mu_x$, $\mu_y$, $\sigma_x$, $\sigma_y$ and $\rho$, respectively, over a single run of ten million
samples of these parameters generated by the Gibbs sampler after allowing for its burn-in phase by
discarding a preceding run of five thousand samples.
The sampling of the fiducial density $f(\rho\,|\,\mu_x,\mu_y,\sigma_x^2,\sigma_y^2,\mathtt{z})$ was
independent from the preceding iterations, while to improve the overall efficiency of the sampling
process, the sampling of the full conditional posterior densities of $\mu_x$, $\mu_y$, $\sigma_x^2$
and $\sigma_y^2$ was based on the Metropolis algorithm.

\begin{figure}[p]
\begin{center}
\makebox[\textwidth]{\includegraphics[width=7.6in]{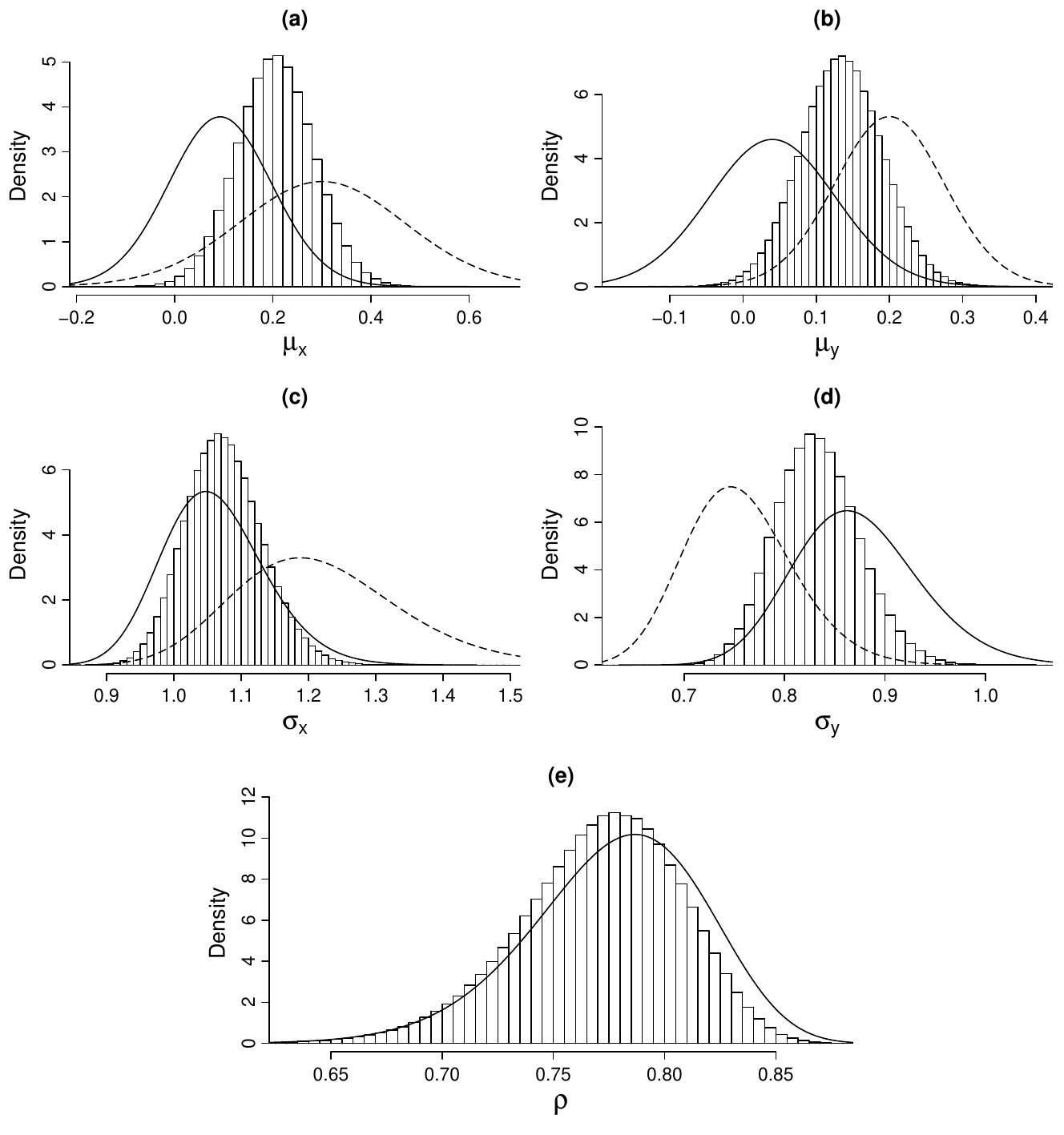}}
\caption{\small{Histograms representing marginal post-data densities of the parameters of a
bivariate normal distribution along with solid curves representing marginal fiducial or confidence
densities of these parameters and dashed curves representing marginal prior densities of $\mu_x$,
$\mu_y$, $\sigma_x$ and~$\sigma_y$}}
\end{center}
\end{figure}

Moreover, the observed data set $\mathtt{z}$ was a typical sample of $n=100$ data points from a
bivariate normal distribution with $\mu_x=0$, $\mu_y=0$, $\sigma_x=1$, $\sigma_y=1$ and $\rho=0.8$.
To give some more details, the sample means of the variables $X$ and $Y$ were 0.0925 and 0.0400,
respectively, the sample standard deviations of these variables were 1.053 and 0.866, respectively,
and the sample correlation between these two variables was 0.780.
In addition, the information required to specify the posterior densities
$p(\mu_x\,|\,\mu_y,\sigma_x^2,\sigma_y^2,\rho,\mathtt{z})$,
$p(\mu_y\,|\,\mu_x,\sigma_x^2,\sigma_y^2,\rho,\mathtt{z})$,
$p(\sigma_x^2\,|\,\mu_x,\mu_y,\sigma_y^2,\rho,\mathtt{z})$ and
$p(\sigma_y^2\,|\,\mu_x,\mu_y,\sigma_x^2,\rho,\mathtt{z})$
was completed by assuming that the values of the constants $\mu'_x$, $\sigma'_x$, $n'_x$, $\mu'_y$,
$\sigma'_y$ and $n'_y$, i.e.\ the constants that control the choice of the prior distributions of
$\mu_x$, $\mu_y$, $\sigma_x^2$ and $\sigma_y^2$ in equations~(\ref{equ4}) and~(\ref{equ5}), have
the following settings: $\mu'_x=0.3$, $\sigma'_x=1.2$, $n'_x=50$, $\mu'_y=0.2$, $\sigma'_y=0.75$
and $n'_y=100$.

In accordance with standard recommendations for evaluating the convergence of Monte Carlo Markov
chains detailed, for example, in Gelman and Rubin~(1992) and Brooks and Roberts~(1998), a
supplementary analysis was carried out in which the Gibbs sampler was run various times from
different starting points and the output of these runs was carefully assessed for convergence using
appropriate diagnostics. This analysis provided no evidence to suggest that the sampler does not
have a limiting distribution, and showed, at the same time, that it would appear to generally
converge quickly to this distribution.

Furthermore, the Gibbs sampling algorithm was run separately with various very distinct fixed
scanning orders of the five model parameters, i.e.\ $\mu_x$, $\mu_y$, $\sigma_x^2$, $\sigma_y^2$
and $\rho$, in accordance with how a single transition of such an algorithm with a fixed scanning
order was defined in Section~4.4 of Bowater~(2018) and without incorporating steps of the
Metropolis sampler into this algorithm as was done in the design of the Gibbs sampling algorithm
just described, i.e.\ without modifying the standard design of a Gibbs sampler.
In doing this, no statistically significant difference was found between the samples of parameter
values aggregated over reasonably long runs of the sampler in using each of the scanning orders
concerned after excluding the burn-in phase of the sampler, e.g.\ between the various correlation
matrices of the parameters and between the various distributions of each individual parameter.
To clarify, by `no statistically significant difference' it is simply meant that none of the
differences between the samples concerned as measured by the summary statistics that were studied
were beyond random chance.
Therefore, taking into account what was discussed in Section~4.5 of Bowater~(2018), it would be
reasonable to conclude that the full conditional densities of the limiting distribution of the
original Gibbs sampler, i.e.\ the one with a uniform random scanning order of the parameters
$\mu_x$, $\mu_y$, $\sigma_x^2$, $\sigma_y^2$ and $\rho$, should be, at the very least, close
approximations to the full conditional densities on which the sampler is based, i.e.\ the posterior
densities $p(\mu_x\,|\,\mu_y,\sigma_x^2,\sigma_y^2,\rho,\mathtt{z})$,
$p(\mu_y\,|\,\mu_x,\sigma_x^2,\sigma_y^2,\rho,\mathtt{z})$,
$p(\sigma_x^2\,|\,\mu_x,\mu_y,\sigma_y^2,\rho,\mathtt{z})$ and
$p(\sigma_y^2\,|\,\mu_x,\mu_y,\sigma_x^2,\rho,\mathtt{z})$ and the fiducial density
$f(\rho\,|\,\mu_x,\mu_y,\sigma_x^2,\sigma_y^2,\mathtt{z})$.

The solid curves overlaid on the histograms in Figures~1(a) and 1(c) are plots of marginal fiducial
densities of the parameters $\mu_x$ and $\sigma_x$, respectively, that were determined according to
the way in which fiducial densities of this type were defined at the end of Section~8 of
Bowater~(2022b) and that correspond to the data set of interest only consisting of the observed
values of the variable $X$, i.e.\ $\{x_i:i=1,2,\ldots,100\}$, while in Figures~1(b) and~1(d), the
solid curves represent, respectively, marginal fiducial densities of $\mu_y$ and $\sigma_y$
determined in the same way except that these densities correspond to the data set of interest only
consisting of the observed values of the variable $Y$, i.e.\ $\{y_i:i=1,2,\ldots,100\}$.
On the other hand, the dashed curves overlaid on the histograms in Figures~1(a) and~1(c) are plots
of the marginal prior densities of the parameters $\mu_x$ and $\sigma_x$, respectively, derived
from the joint prior density of $\mu_x$ and $\sigma_x$ that is defined by the conditional prior
densities of $\mu_x$ and $\sigma_x$ specified in equations~(\ref{equ4}) and~(\ref{equ5}).
Similarly, in Figures~1(b) and~1(d), the dashed curves represent, respectively, the marginal prior
densities of $\mu_y$ and $\sigma_y$ derived from the joint prior density of $\mu_y$ and $\sigma_y$
that is defined by the conditional prior densities of $\mu_y$ and $\sigma_y$ specified in
equations~(\ref{equ4}) and~(\ref{equ5}).

Finally, the solid curve overlaid on the histogram in Figure~1(e) is a plot of a confidence density
function for the correlation $\rho$. In general, a density function of this type corresponds to a
set of confidence intervals over which the coverage probability for the parameter concerned
gradually varies, see for example Efron~(1993) for further clarification.
More specifically, for the plot being considered, these confidence intervals for $\rho$ were
constructed on the basis of summarising the data set $\mathtt{z}$ by the sample correlation
coefficient $r$, and then assuming that the Fisher transformation of this coefficient, i.e.\ the
transformation $\tanh^{-1}(r)$, has a normal sampling distribution with mean $\tanh^{-1}(\rho)$ and
variance $1/(n-3)$, which is a standard method that is used in practice to form confidence
intervals for the correlation $\rho$.

By comparing the histograms in Figures~1(a) to~1(e) with the curves that have been overlaid on
them, it can be seen that the forms of the marginal post-data densities of $\mu_x$, $\mu_y$,
$\sigma_x$, $\sigma_y$ and $\rho$ that are represented by these histograms are consistent with what
we would have intuitively expected given the way in which pre-data knowledge about the parameters
$\mu_x$, $\mu_y$, $\sigma_x$ and $\sigma_y$ and a lack of pre-data knowledge about the correlation
$\rho$ has been taken into account by the method of inference that has been applied.

\end{document}